\documentclass[a4paper]{eccomas_paper-2024}
\usepackage{graphicx}
\usepackage{subfig}

\usepackage{hyperref}

\usepackage{graphicx}
\usepackage{amsmath}
\usepackage{amsfonts}
\usepackage{amssymb}

\title{DIFFERENTIALLY HEATED TURBULENT CHANNEL FLOW TWO-POINT CORRELATIONS}

\author{Marina Garcia-Berenguer$^{1*}$, Lucas Gasparino$^{2}$, Oriol Lehmkuhl$^{2}$ and Ivette Rodriguez$^{1}$}

\heading{Marina Garcia-Berenguer, Lucas Gasparino, Oriol Lehmkuhl and Ivette Rodriguez}

\address{$^{1}$ Turbulence and Aerodynamics Research Group (TUAREG)\\
Universitat Politècnica de Catalunya (UPC), 08222, Terrassa (Spain)
\and
$^{2}$ Large-scale Computational Fluid Dynamics\\
Barcelona Supercomputing Center (BSC), 08034, Barcelona (Spain)
\and
$^{*}$ marina.garcia.berenguer@upc.edu}


\abstract{This study analyzes the behavior of a differentially heated channel flow by means of a direct numerical simulations (DNS) with variable thermophysical properties under low-speed conditions focusing on the impact of the temperature gradient on the turbulence structures near the channel walls. The simulations were conducted at a mean friction Reynolds number of $Re_{\tau m} = 400$ with a temperature ratio between the walls of $T_{hot}/T_{cold} = 2$. Results show significant differences between the hot and cold walls that lead to an increased heat transfer at the hot wall and a higher turbulent production in the cold wall.}

\begin{document}
\thispagestyle{empty}

\section{INTRODUCTION}

Channel flow numerical simulations have been used to study the behavior of wall-bounded flows both in engineering applications and research. The complexity of these flows increases when the channel is heated, especially if its thermophysical properties (density, viscosity, and thermal conductivity) are assumed to be variable and a temperature gradient is added between the walls.

Early studies added the temperature as a passive scalar to an incompressible flow \cite{KimMoinMoser1987, Kawamura2001, Kawamura2004} to analyze the effects of the flow at different friction Reynolds numbers, different Prandtl numbers, and different heat flux conditions at the wall. 
The friction Reynolds number is defined as $Re_\tau = \rho_w \delta u_\tau / \mu_w$, where $u_\tau=\sqrt{\tau_w/\rho_w}$ is the friction velocity, $\delta$ is the channel half-height, $\tau_w$ is the wall shear stress, $\rho_w$ and $\mu_w$ are the density and the dynamic viscosity at the wall, respectively. The Prandtl number, defined as $Pr=\mu c_p/\lambda$ where $c_p$ is the specific heat and $\lambda$ is the thermal conductivity, is set as a constant.

Few studies have been carried out on channels with an imposed temperature at the walls and with variable thermophysical properties and most of them focused on high-speed flows. Huang et al. \cite{Huang1995} analyze the different models and scalings at high-speed flows. Furthermore,  direct numerical simulations (DNS) of compressible channel flows were conducted by Modesti and Pirozzoli \cite{Modesti2016} to study the flow structures at different friction Reynolds numbers and bulk Mach numbers (between 1.5 and 3), focusing on compressibility effects and turbulence structure variations.

In low-speed flows, the research is mainly about differentially heated walls. Some large-eddy simulations (LES) have been performed reaching a temperature ratio between the walls up to $T_{cold}/T_{hot} = 9$ but for low mean friction Reynolds number $Re_{\tau m}= 0.5(Re_{\tau cold} + Re_{\tau hot})=180$ \cite{Lessani}. Other studies have conducted LES at higher friction Reynolds number, $Re_\tau=395$, but at low-temperature ratios and using the low-Mach approximation, therefore neglecting the possible compressible effects due to thermo-dependant properties \cite{Serra}.

Additionally, Patel et al. \cite{Patel2016} investigated the impact of a low-temperature gradient between the walls using different constitutive relations for the thermophysical properties at low Reynolds, focusing on the effect on the velocity streaks and the turbulent structures. 

This type of study has increased in recent years due to the interest in transcritical and supercritical fluids. For example, Doehring and Adams \cite{Doehring2018} conducted LES of transcritical channel flow with real gases, with a temperature ratio of 2.2 and a friction Reynolds number of 400.

Toutant and Bataille \cite{Toutant2013} performed DNS of a low-Mach channel flow with variable thermophysical properties at mean friction Reynolds number $Re_{\tau m} = 400$ and temperature ratio $T_{hot}/T_{cold}=2$ between the walls. They studied the flow with different scalings for the velocity and temperature profiles. However, the effect of the asymmetrical heat on the turbulent structures was not discussed.

The present work aims to go beyond and study the behavior of these differentially heated channel flows focusing on the turbulent structures of the walls through the two-point velocity correlations. A direct numerical simulation with variable thermophysical properties and at low speed has been conducted at mean friction Reynolds number of $Re_{\tau m}= 400$ and at temperature ratio difference between the walls, $T_{hot}/T_{cold}$, of 2.

\section{NUMERICAL METHODOLOGY}
The full compressible Navier-Stokes equations were used to simulate the heated channel flow and capture all the variations of the thermophysical properties. However, the Mach number has been kept below 0.2 to ensure a low-speed flow.

\begin{equation}
    \dfrac{\partial \rho}{\partial t} + \dfrac{\partial \rho u_i}{\partial x_i} = 0
    \label{eq:1}
\end{equation}

\begin{equation}
    \dfrac{\partial \rho u_i}{\partial t} + \dfrac{\partial \rho u_i u_j}{\partial x_j} = - \dfrac{\partial p}{\partial x_i} + \dfrac{\partial \tau_{ij}}{\partial x_j} + f_i
    \label{eq:2}
\end{equation}

\begin{equation}
    \dfrac{\partial \rho E}{\partial t} + \dfrac{\partial \rho u_i(E+p)}{\partial x_i} = \dfrac{\partial u_i \tau_{ij}}{\partial x_i} + \dfrac{\partial q_i}{\partial x_i} + s
    \label{eq:3}
\end{equation}

 In these equations, $x_i=(x,y,z)$ are the spatial coordinates, $t$ is the time, $u_i = (u, v, w)$ are the velocity components, $\rho$ is the density, $p$ is the dynamic pressure and $f_i$ and $s$ are the source terms for momentum and energy conservation, respectively.
 The stress tensor, $\tau_{ij}$, is defined as $\tau_{ij}=2\mu\left(\dfrac{1}{2}\left(\dfrac{\partial u_j}{\partial x_i}+\dfrac{\partial u_i}{\partial x_j}\right)-\dfrac{1}{3}\dfrac{\partial u_k}{\partial x_k}\delta_{ij}\right)$, where $\mu$ is the dynamic viscosity. 
 
 The energy per unit mass is defined as $E=\rho\left(\dfrac{1}{2}u_i u_i + e\right)$, with $e$ as the internal energy. Finally, the heat flux is defined as $q_i=-\lambda \dfrac{\partial T}{\partial x_i}$, where $\lambda$ is the thermal conductivity.

To solve these equations and the thermo-dependent variables, the ideal gas approach was used.
 \begin{equation}
    P_0 = \rho R_g T
    \label{eq:state}
\end{equation}

Air was set as the working fluid with a gas constant of $R_g=286.86$ $J \ kg^{-1} \ K^{-1}$ and a thermodynamic pressure of $P_0=101325$ $Pa$.

The dynamic viscosity $\mu$ and the thermal conductivity $\lambda$ were computed using Sutherland's law \cite{Sutherland}, where the Prandtl number was fixed at $Pr=0.71$ and the specific heat capacity for the air is $c_p=1004$ $J \ kg^{-1} \ K^{-1}$.

\begin{equation}
    \mu(T) = 1.458 \times 10^{-6} \frac{T^{3/2}}{T+110.4}
\label{eq:SLmu}
\end{equation}

\begin{equation}
    \lambda (T) = \frac{\mu c_p}{Pr}
\label{eq:SLlambda}
\end{equation}

The simulations were conducted using {\it sod2d}, an in-house code developed by the Barcelona Supercomputing Center \cite{sod2dPaper, sod2d}. The code is designed for high-fidelity scale-resolving simulations, such as large eddy simulations (LES) and direct numerical simulations (DNS), based on low-dissipation high-order spectral elements. The algorithm combines Galerkin's spectral-method model for continuous finite elements with a modified version of Guermond's entropy viscosity stabilization, which is tailored to work seamlessly with the spectral elements approach.

\begin{figure}[!h]
\centering
  \includegraphics[width=0.55\textwidth]{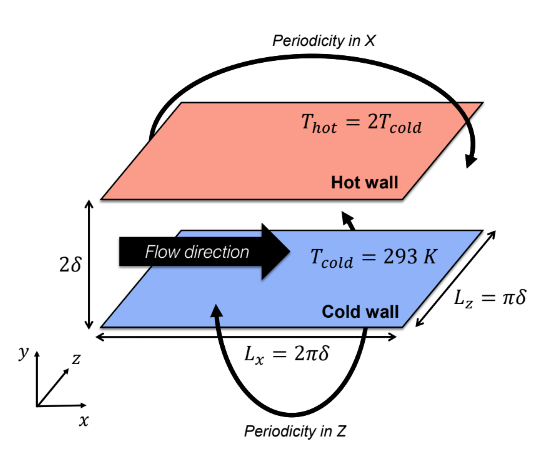}
\caption{Case configuration and domain schematics.}
\label{fig:dom}
\end{figure}

To overcome the aliasing effects resulting from the reduced order integration due to using SEM, the scheme employs an operator split for the convective terms. Additionally, the spectral-type elements implemented in {\it sod2d} employ the Lobatto-Gauss-Legendre (LGL) quadrature, providing computational advantages in terms of cost compared to conventional models, particularly when hexahedron-shaped elements are employed.

For the time-advancing algorithm, an explicit fourth-order Runge-Kutta method is utilized.

The computational domain for the heated channel is set with two periodic directions in the $x$ and $z$ axis, as seen in figure \ref{fig:dom}. The dimensions of the domain are $L_x = 2\pi\delta$, $L_y=2\delta$, $L_z=\pi\delta$ \cite{delAlamo2004}, where $\delta=0.003$ is the half height of the channel. The temperature is fixed and constant at the no-slip walls. The temperature at the cold wall is set at $T_{cold}=293$ $K$ and, to ensure a temperature ratio of $T_{hot}/T_{cold} = 2$, the hot wall is set at $T_{hot}=586$ $K$.

The fourth-order hexahedral mesh has more than 50 million grid points that allow the flow to be fully resolved. The grid spacings, which can be seen in the table \ref{tab:grid} for both walls, are computed using each wall specific friction velocity $u_\tau$ ($\Delta x^+_i = \Delta x_i \rho  u_\tau/ \mu$).

Finally, the simulation statistics were integrated for 13.93 mean diffusion time units at the statistical steady state, where the mean diffusion time is computed with the mean friction velocity between the walls $t_d = \delta/u_{\tau m}$ (being $u_{\tau m} = 0.5(u_{\tau cold} + u_{\tau hot}$). For each specific wall the simulation time is calculated with their specific friction velocity, as shown in table \ref{tab:grid}.

\section{RESULTS}

The averaged simulation results are analyzed to understand the behavior of the differentially heated channel flow at the different walls. These results were time-averaged using the Favre approach \cite{Favre}, which considers the variable density of the channel. All the results are normalized by the local wall friction velocity $u_\tau$ and by the local friction temperature $T_\tau$, defined as $T_\tau = q_w/(c_p \rho_w u_\tau)$.

\begin{table}[!h]
    \centering
    \begin{tabular}{ccccccc}
        Case wall&  $Re_\tau$ & $u_\tau$ & $\Delta x^+$ & $\Delta y^+$ & $\Delta z^+$ & $T_s/t_d$\\ \hline
         Cold wall& 560.08 & 2.89 & 9.96 & 0.93-4.85 & 4.20 &12.20 \\
         Hot wall & 236.32 & 3.85 & 4.98 & 0.87-2.05 & 2.10 & 16.23\\
    \end{tabular}
    \caption{Simulation parameters at each wall.}
    \label{tab:grid}
\end{table}

The results of each wall are summarized in the table \ref{tab:grid}. The friction Reynolds number obtained at each wall varies from 560.08 to 236.32 due to the temperature gradient between the walls. This means that there is a different turbulent regime at each half of the channel, making it possible to study a more turbulent flow and a more laminar flow with the same configuration.
The mean friction Reynolds number obtained from the simulation is $Re_{\tau m} =398.20$ with a maximum local Mach number of 0.16.

The behavior of the differentially heated channel flow varies significantly from that of the incompressible flow. The effects of these temperature differences are evident in the statistical and structural analysis of the flow.

\begin{figure}[]
\centering
\subfloat[Velocity profile] {\includegraphics[width=0.45\textwidth]{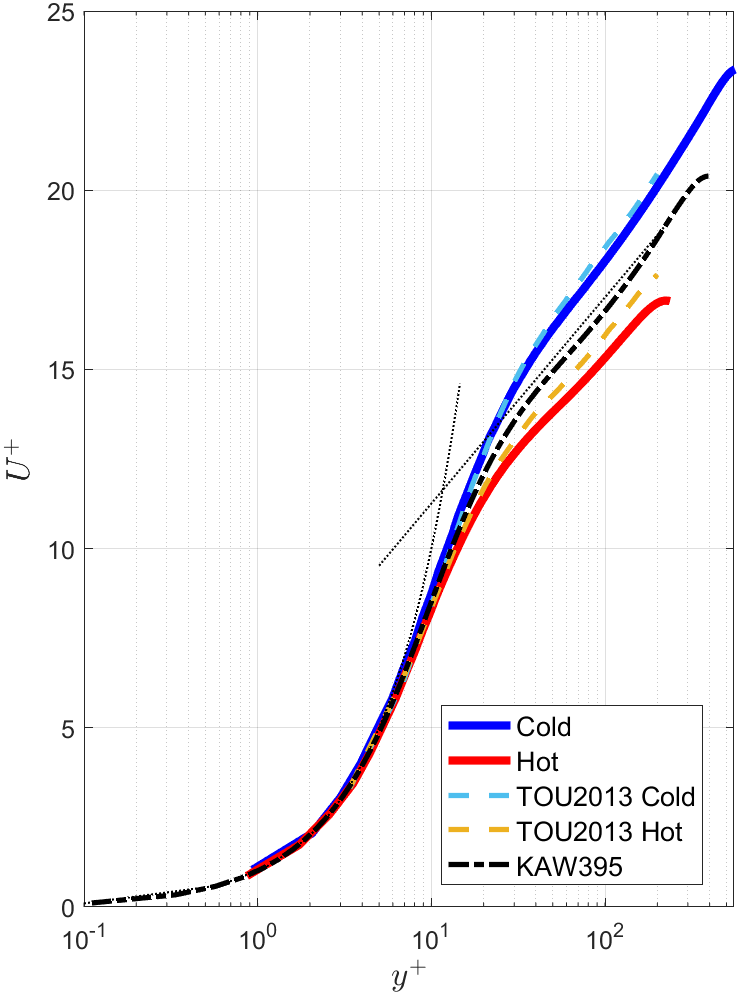}}\hfill
\subfloat[Temperature profile] {\includegraphics[width=0.42\textwidth]{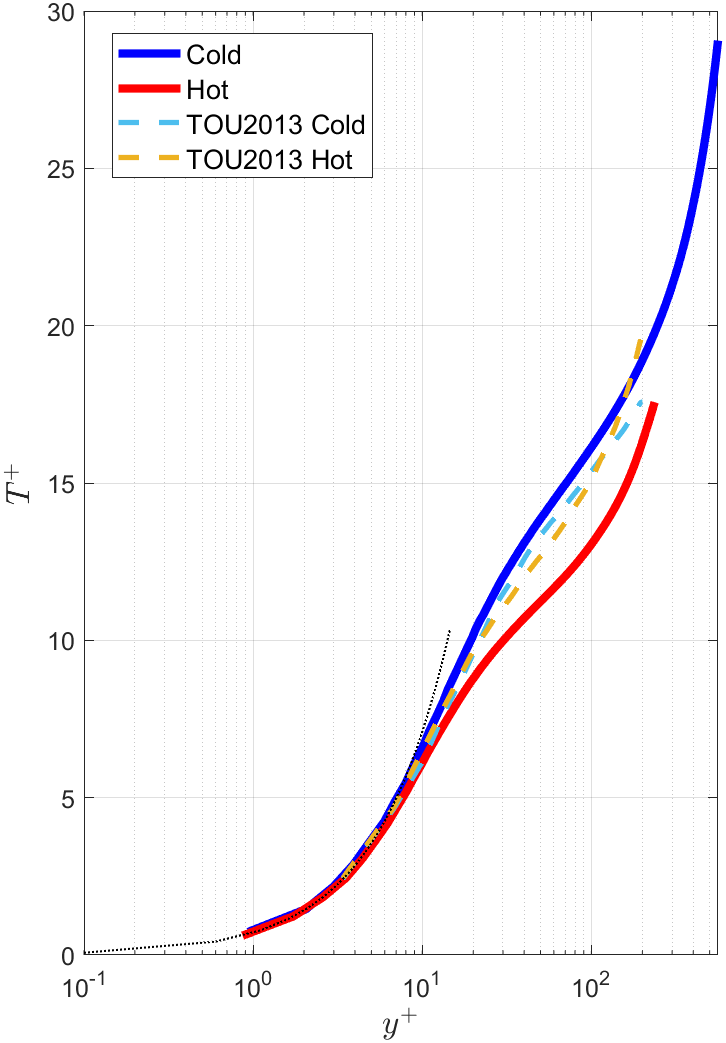}}\hfill
\caption{Mean profiles, where the cold wall is in blue and the hot wall in red, compared with the incompressible case in black \cite{Kawamura2001} and with Toutand and Bataille \cite{Toutant2013} in dashed lines.}
\label{fig:U}
\end{figure}

Figure \ref{fig:U} shows the mean velocity and the mean temperature profiles for the cold and hot walls, compared with the results from Toutant and Bataille \cite{Toutant2013} and the incompressible case computed at the same friction Reynolds number, $Re_\tau = 395$ \cite{Kawamura2001}.  The results show good agreement with the simulations by Toutant and Bataille \cite{Toutant2013}, validating them despite the temperature deviations caused by the different case approaches. The difference between the velocity profiles and the classical log-law of the incompressible case is due to the temperature gradient.

\newpage 

The mean temperature inside the channel decreases due to the temperature gradient at the walls, which is steeper at the cold wall. The increase in cold friction Reynolds number indicates an increase of the turbulent energy near that wall, in contrast to what occurs at the hot wall. This phenomenon can be explained by the higher viscosity at the hot wall, which leads to higher dissipation.

\begin{figure}[!h]
\centering
  \subfloat[Streamwise velocity $u_{rms}^{\prime \prime}$] {\includegraphics[width=0.5\textwidth]{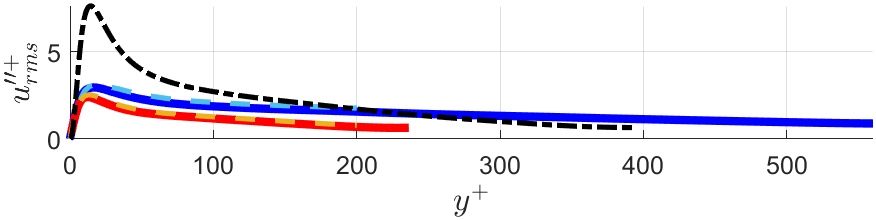}}\hfill
  \subfloat[Normal velocity $v_{rms}^{\prime \prime}$] {\includegraphics[width=0.5\textwidth]{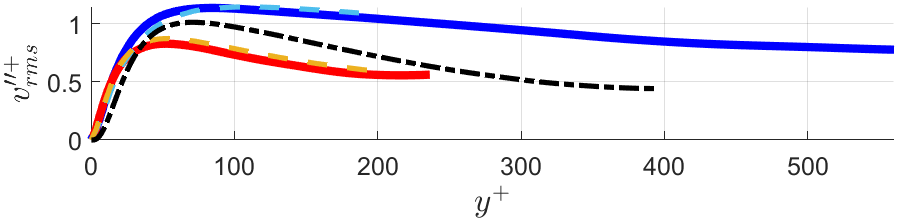}}\hfill
  \subfloat[Spanwise velocity $w_{rms}^{\prime \prime}$] {\includegraphics[width=0.5\textwidth]{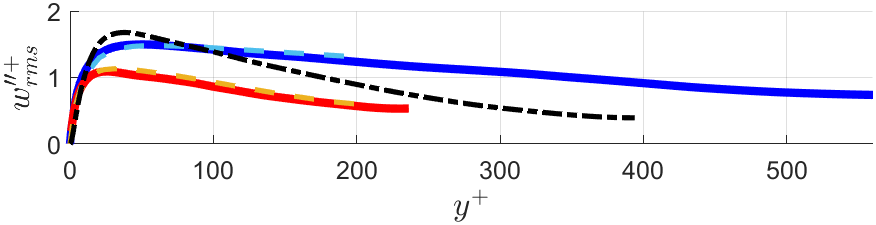}}\hfill
  \subfloat[Shear stress $u^{\prime \prime}v^{\prime \prime}$] {\includegraphics[width=0.5\textwidth]{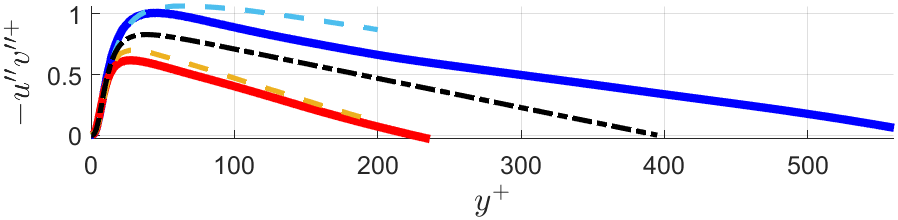}}
\caption{Reynolds stresses fluctuations where the cold wall is represented in blue and the hot wall in red. All compared with the incompressible case in black \cite{Kawamura2001} and with Toutand and Bataille \cite{Toutant2013} in dashed lines.}
\label{fig:uu}
\end{figure}

The Reynolds stresses, in figure \ref{fig:uu}, show the differences between the cold wall, the hot wall, and the incompressible case at the same friction Reynolds number.

Applying a thermal gradient between the walls has big effects. First of all, the streamwise velocity fluctuations are significantly less important inside the viscous sublayer of the channel than in the incompressible case. However, in the center of the channel, the normal and the spanwise velocity fluctuations are more relevant, especially at the cold wall. This is due to the density difference between the walls that increases the turbulent mixing of the flow in the wall direction. 

The shear stresses present a similar pattern, proving that the cold wall has a more turbulent behavior which aligns with the fact that the friction Reynolds number is higher in that wall.

The temperature fluctuations increase in the outer viscous sub-layer of the channel, enhancing the heat transfer between the walls. In the inner viscous sub-layer, the streamwise heat flux presents a peak, which is higher at the cold wall, indicating more local turbulence mixing near the wall, especially near the cold one.

The two-point correlations of the streamwise velocity were computed to understand the flow behavior near the wall and its turbulent structures. Therefore, the correlations have been computed centered near the walls of the channel, at $y_{ref}/\delta \approx 0.1$, using the equation \ref{eq}. The specific reference heights for each wall are shown in table \ref{tab:corr}. 

\begin{equation}\label{eq}
    R_{ij}^x(r_x, r_y, y_{ref}) = \langle \langle u_i^{\prime\prime}(x, y_{ref}, z, t) \ u_j^{\prime\prime}(x+r_x, y_{ref}+r_y, z, t) \rangle\rangle_{t, z} 
\end{equation}

where $\vec{r} = (r_x, r_y, r_z)$ is the space vector separating the two points of the domain at the same time, the $\langle \cdot \rangle$ is the ensemble averaging operator, and the $\langle \cdot \rangle_{t,z}$ represents the averaging over time and the $z$ plane.

\begin{table}[!h]
    \centering
    \begin{tabular}{ccc}
         Case wall & $y_{ref}/\delta$ & $y_{ref}^+$\\ \hline
        Cold wall & 0.101 & 56.57\\ 
        Hot wall & 1.902 & 448.63\\
    \end{tabular}
    \caption{Correlations reference height $y_{ref}$ at each wall.}
    \label{tab:corr}
\end{table}

\begin{figure}[]
\centering
  \includegraphics[width=0.8\textwidth]{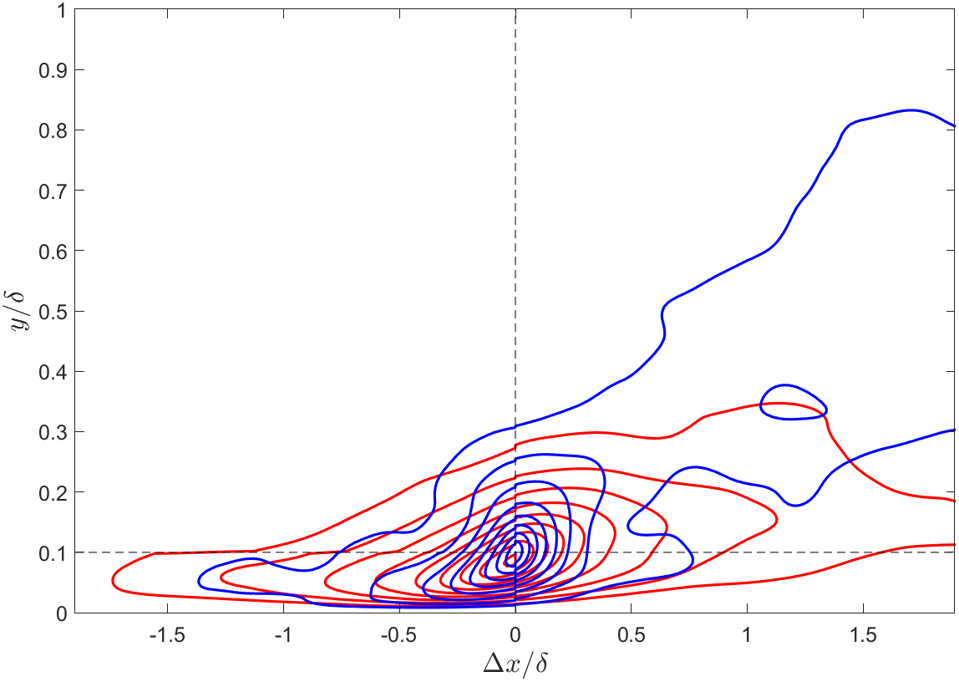}

\caption{Streamwise velocity $xy$ two-point correlations. The cold wall correlations, in blue, are computed at $y_{ref}/\delta = 0.1$, and the hot wall correlations, in red, at $y_{ref}/\delta = 1.9$.}
\label{fig:corr2D}
\end{figure}

Figure \ref{fig:corr2D} shows the streamwise velocity correlations centered at the reference height. Here, the cold wall presents correlated structures that evolve towards the center of the channel suggesting high turbulent activity. This results in a higher production of small structures near the cold wall. This turbulent production is then transported in the wall direction due to the density gradient making the structures more important in the outer layer of the cold wall of the channel.

However, the hot wall presents even more correlated and elongated structures compared to the cold wall, implying fewer structures and thus less turbulent mixing. The structures near the hot wall are larger and extremely streamwise oriented because the temperature gradient increases the heat transfer efficiency at the hot wall, making the flow to behave almost laminar.

\newpage 
In the spanwise direction, the $zy$ correlations (not shown here) present a similar behavior. The cold wall presents larger and more dispersed structures while the hot wall presents more compact and smaller ones. Remarkably, these correlations are not symmetrical, as observed in the incompressible case. The temperature gradient forces a recirculation of the flow between the cold and the hot wall where the high turbulent energy produced at the cold wall is transported to the hot wall and dissipated there.
Finally, the hot wall, which presents smaller structures in the $zy$ plane, experiences higher viscous and thermal dissipation than the cold wall, reducing its thermal layer.

\section{CONCLUSIONS}

In this study, a direct numerical simulation (DNS) was conducted to analyze the effects of differential heating on channel flows with variable thermophysical properties. The results show that the temperature gradient between the walls has a significant effect on the turbulence structures of the channel. 

The analysis of the statistics shows that there is more flow mixing and heat transfer between the walls.
At the cold wall, the higher local turbulent production increases the flux mixing and the thermal transport. On the other hand, the hot wall presents a quasi-laminar flow due to viscous dissipation at the wall that increase the efficiency of the heat transfer.

Studying the structures near the walls, the cold wall presented a higher level of turbulence with smaller, more dispersed, and numerous structures, while the hot wall showed larger, elongated, and more streamwise-oriented structures.

\newpage 
The recirculation of the flow, from the cold to the hot wall, forced by the temperature gradient breaks the symmetry of the $zy$ two-point correlations. This asymmetry reinforces the different behavior of each wall.

In short, this work explains how a temperature gradient affects the turbulent structures in wall-bounded flows. The channel, at $Re_{\tau m}$ and $T_{hot}/T{cold}=2$, proves that the compressibility effects of the thermo-dependent variables are not neglectable and should be further studied.


\section*{Acknowledgments}
This work has been partially financially supported by the \textit{Ministerio de Econom\'ia y Competitividad, Secretar\'ia de Estado de Investigaci\'on, Desarrollo e Innovaci\'on}, Spain (ref. PID2020-116937RB-C21 and PID2020-116937RB-C22). The corresponding first author gratefully acknowledges the Universitat Politècnica de
Catalunya and the Banco Santander for the financial support of her predoctoral grant FPI-UPC.


\bibliographystyle{unsrt}
\bibliography{biblio}


\end{document}